\documentclass[]{aastex631}

\submitjournal{APJL}

\shorttitle{Solar Chromospheric differential rotation}
\shortauthors{Kharayat et al.}
\graphicspath{{./}{figures/}}
\usepackage{graphicx}

\begin{document}

\title{Equator to Pole Solar Chromospheric Differential Rotation using Ca-K Features Derived from Kodaikanal Data}

\author[0000-0003-3522-3135]{Hema Kharayat}
\affiliation{Department of Physics \\
M. L. K. P. G. College \\
Balrampur, 271201, Uttar Pradesh, India}
\affiliation{Indian Institute of Astrophysics \\
Bangalore, 560034, India }

\author{Jagdev Singh}
\affiliation{Indian Institute of Astrophysics \\
Bangalore, 560034, India }


\author[0000-0002-4339-8608]{Muthu Priyal}
\affiliation{Indian Institute of Astrophysics \\
Bangalore, 560034, India }

\author[0000-0003-2165-3388]{B. Ravindra}
\affiliation{Indian Institute of Astrophysics \\
Bangalore, 560034, India }





\begin{abstract}

Differential rotation is one of the basic characteristics of the Sun, and it plays an important role in generating the magnetic fields and its activities. We investigated rotation rate using chromospheric features such as plages, enhanced network (EN), active network (AN), and quiet network (QN) separately (for the first time). The digitized Ca-K images from Kodaikanal Observatory for 1907--1996 are used to study rotation over 0--80  degrees latitudes at an interval of 10$^{\circ}$. We find that plages and all types of networks exhibit the differential rotation of the chromosphere. Furthermore, the rotation rate shows a decreasing pattern as one move from the equator to the higher polar latitudes for all the features used in the study. At the equator the rotation rate (rotation period) is obtained to be $\sim$13.98 deg day$^{-1}$ (25.74 days), $\sim$13.91 deg day$^{-1}$ (25.88 days), $\sim$13.99 deg day$^{-1}$ (25.74 days), and $\sim$14.11 deg day$^{-1}$ (25.51 days) for plage, EN, AN, and QN areas, respectively. By analyzing how the area of chromospheric features varies over time,  we can effectively map the Sun's rotation rate at all latitudes, including the polar regions. Interestingly, both plages and small-scale networks  exhibit similar differential rotation rate. This suggests these features likely rooted at the same layer below the visible surface of the Sun. Therefore, the long-term Ca-K data is very useful to study the solar rotation rate at all latitudes including the polar regions. 


\end{abstract}



\section{Introduction} \label{sec:intro}
Solar differential rotation is a topic of investigation among researchers due to its close connection with solar dynamo theory \citep{Babcock1961ApJ, Dikpati2006ApJ, Chu2010SoPh, Javaraiah2020SoPh, Hotta2021NatAs}. The interaction of the Sun’s differential rotation with the magnetic field plays a fundamental role in generating all solar activities \citep{Javaraiah2005ApJ}. Therefore, studying solar differential rotation is essential for better understanding the physical mechanisms behind the various solar activities and cycles.

Solar differential rotation has been studied for a long time using various forms of solar activity indexes. Sunspots are one of the oldest recorded solar parameters and have been widely used for the measurement of the rotation rate of the solar atmosphere \citep[and references therein]{Lustig1983AA, Gupta1999SoPh, Javaraiah2005SoPh}. \cite{Ruzdjak2017SoPh} analysed sunspot groups to find a relationship between the solar rotation and activity. They found that the Sun rotates more differentially at the minimum than at the maximum of activity. By analysing the sunspot data from the Royal Greenwich Observatory and the USAF/NOAA, \cite{Zhang2013AA} found evidence of an anticorrelation of the rotation in the two hemispheres and oscillatory behavior of the asymmetry at a period of 80--90 years. They also found that the north-south asymmetry of solar rotation have an inverse relationship with the area of large sunspots. \cite{Jha2021SoPh} measured the solar rotation profile using the white-light sunspot data from Kodaikanal Observatory. They found no variations in rotation rate between activity extremes, i.e., solar maxima and minima. To study the coronal rotation, \cite{Insley1995SoPh} considered coronal holes as tracers of the differential rotation. They noticed that the mid-latitude corona rotates more rigidly than the photosphere, but still exhibits significant differential rotation. \cite{Vats2001ApJ} derived the solar coronal rotation from the disk-integrated simultaneous daily measurements of solar flux at eleven radio frequencies in the range of 275--2800 MHz. They found that the sidereal rotation period decreases with the altitude. \cite{Wohl2010AA} determined the solar rotation velocity by tracing small bright coronal structures in SOHO-EIT images. They found that it has a rotation velocity similar to those obtained by small photospheric magnetic features. The differential rotation rate of the solar corona at low-latitudes is found to be consistent with the rotational profile of the near-surface convective zone of the Sun \citep{Mancuso2020AA}. By analysing the observations from the Atmospheric Imaging Assembly (AIA) telescope on board the Solar Dynamics Observatory (SDO) at different wavelengths, \cite{Sharma2020MNRAS} found that the sidereal rotation periods of different coronal layers decrease with increasing temperature (or height).

The Ca II K line images serve as an excellent diagnostic tool to investigate the rotation of the solar chromosphere. A wide range of previous studies are available for the chromospheric differential rotation \citep[and references therein]{Livingston1969SoPh, Belvedere1977ApJ, Antonucci1979SoPhl,Antonucci1979SoPh, Singh1985SoPh,Ternullo1986SoPh, Ternullo1987SoPh, Li2020ApJ, Xu2020ApJ, Wan2022ApJ, Li2024MNRAS}.  Using the total plage area integrated over the entire visible hemisphere of the Sun, \cite{Singh1985SoPh} found a dominant periodicity of 7 years in rotation rate. \cite{Bertello2020Apj} analyzed the full-disk Ca II K images obtained from Mount Wilson Observatory. They found that temporal variations in full-disk chromospheric activity show the signature of the 11 year solar cycle. Also, they noticed that there is no indication of any detectable periodicity in the temporal behavior of the orthogonalized rotation rate coefﬁcients. \cite{Schroeter1975SoPh} and \cite{Belvedere1977ApJ} investigated differential rotation of Ca II Networks and Plages, respectively. They found that the chromosphere rotates 5$\%$ faster than the photosphere. \cite{Li2020ApJ} and \cite{Xu2020ApJ} utilized data in He I and Mg II lines and reported a faster rotation of the chromosphere. In an attempt to measure the chromospheric rotation rate, \cite{Wan2022SoPh} utilize Ca II K filaments and suggested faster rotation of chromosphere. Recently, \cite{Mishra2024ApJ} utilized Ca II K data obtained from Kodaikanal Solar Observatory and reported that the chromospheric plages exhibit an equatorial rotation rate 1.59$\%$ faster than the photosphere. At the chromospheric layer, the short-lived features of Ca II K rotate at the same rate as the chromospheric plasma \citep{Antonucci1979SoPhl}. Recently, \cite{Wan2023MNRAS} analyzed synoptic maps of Ca II K-normalized intensity to investigate the long-term variation of the quiet chromospheric differential rotation. They found that the rotation rate is smaller for the quiet chromosphere than for the chromosphere on the whole.
 
\cite{Hathaway2011ApJ} measure differential rotation of the small magnetic elements using the data from Michelson Doppler Interferometer (MDI) instrument on board Solar and Heliospheric Observatory (SOHO) spacecraft \citep{Scherrer1995SoPh}. They found that the differential rotation vary  systematically over the solar cycle and the differential rotation is weaker at maximum than at minimum. By tracking the motions of individual magnetic features from a single month of observations from a Helioseismic and Magnetic Imager (HMI) aboard SDO, \cite{Lamb2017ApJ} has arrived at high-precision measurements of solar rotation. \cite{Imada2018ApJ} studied the dependence of the surface flow velocity on the magnetic field strength. They found that magnetic elements with a strong (active region remnants) and weak (solar magnetic networks) magnetic field show a faster and slower rotation speed, respectively. \cite{Hathaway2022FrASS} measured differential rotation and meridional flow in the Sun's surface shear layer by tracking the motions of the magnetic network using magnetic pattern tracking on magnetograms from the MDI on board SOHO spacecraft and from HMI on board SDO during 1996-2022. They found that  both, differential rotation and meridional flow, vary in strength with depth. The rotation rate increases inward while the meridional flow weakens inward.

In this way, many authors have studied the solar rotation rate by using the sunspot numbers and others using the whole disk solar chromospheric and coronal features but not the small-scale networks. In a study of differential rotation by tracking the motions of the magnetic network, \cite{Hathaway2022FrASS} found slower flows on the poleward sides of the active latitudes and faster flows equatorward. In our study, we also found a decreasing trend in the differential rotation, of both plages and small-scale networks, towards the poles.  Further, most of the investigations made in the past have considered only up to mid-latitudes (up to $\pm$ 50$^{\circ}$) for the study of differential rotation. Recently, \cite{Singh2021ApJ} have generated a uniform time series of Ca-K images using the Equal Contrast technique (ECT), which permitted the extraction of the reliable values of areas of small-scale features in daily available Ca-K data. The uniqueness of the present work lies in the study of solar differential rotation by using various small-scale solar chromospheric features, viz plage, Enhanced Network (EN), Active Network (AN), and Quiet Network (QN) areas. Furthermore, with this data, we are fortunate enough to extend the study of solar rotation profile up to higher polar latitudes (up to $\pm$ 80$^{\circ}$).

In this paper, we present the long-term measurements (1907–1996) of average solar rotation rate as a function of heliographic latitude with the help of the fast Fourier transform method. We utilize the digitized data of plage area and small-scale networks obtained from the Kodaikanal Observatory. The paper is structured as follows. Section \ref{sec:data} provides the information about the data and method used for the present study. In Section \ref{sec:result}, we discuss the results of the investigation. The paper ends with conclusions about the results obtained.

\section{data and Method} \label{sec:data}

\begin{figure*}[!ht]
\gridline{\fig{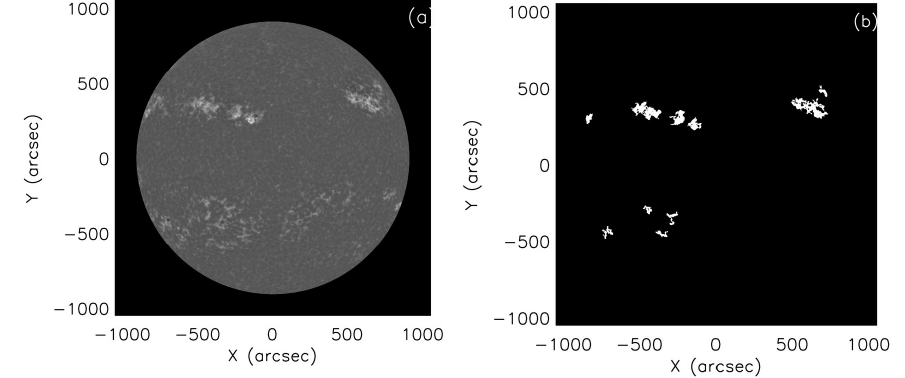}{0.85\textwidth}{}
          }
\gridline{\fig{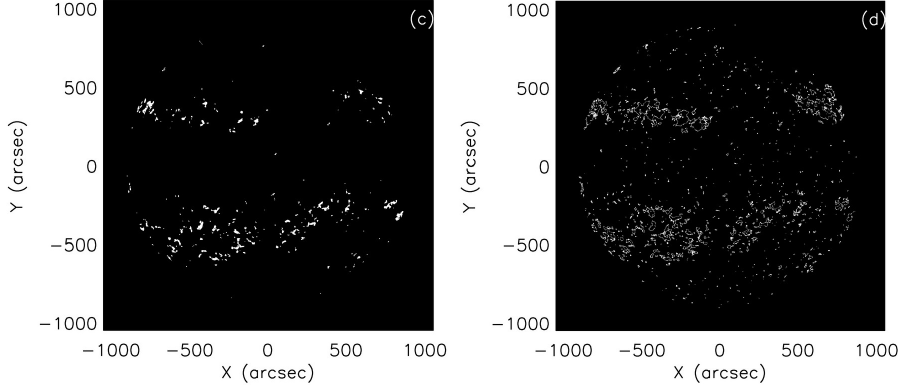}{0.85\textwidth}{}
          }
\gridline{\fig{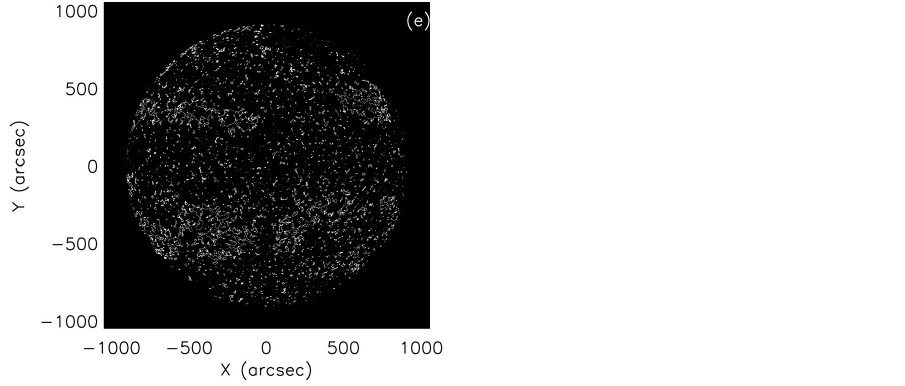}{0.85\textwidth}{}}
\caption{Panel (a) of the figure shows the analyzed Ca-K image taken on 1936 April 11, 07:46:00 UT. Panels (b), (c), (d), and (e) show the identified plages, EN, AN, and QN in the binary format, respectively.}
\label{fig:fig_1}
\end{figure*}
\begin{figure}[!ht]
\plotone{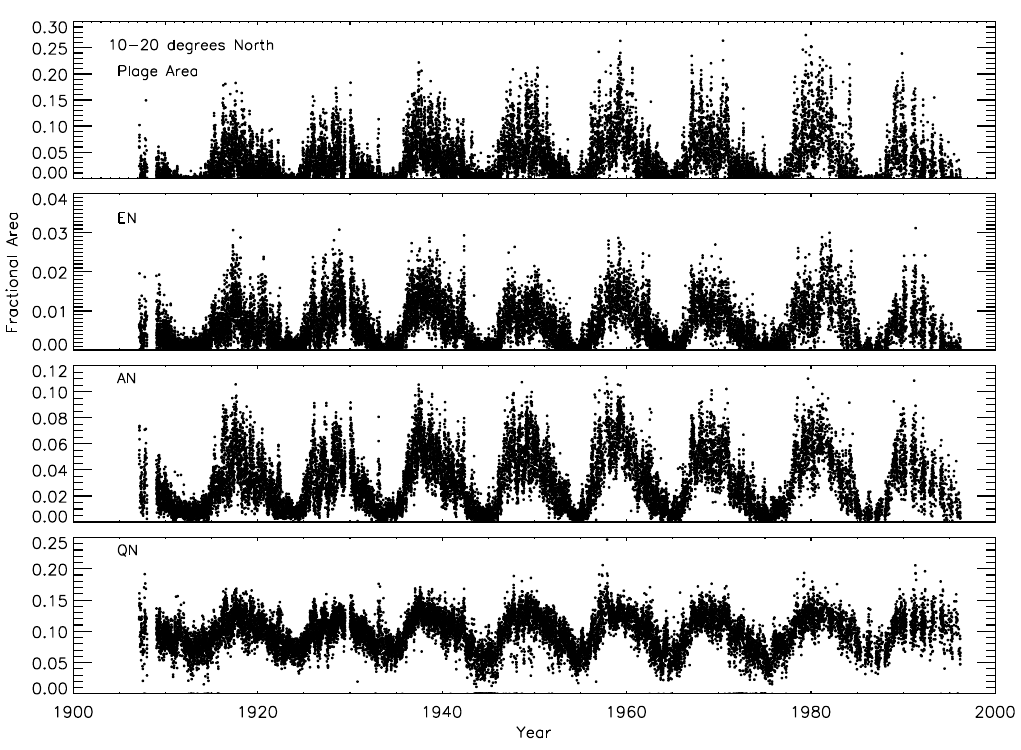}
\caption{(Top to bottom) Distribution of daily fractional area of plages, EN, AN, and QN for the period 1907-1996 for 10-20$^{\circ}$ latitudinal belt of northern hemisphere.} 
\label{fig:fig_frac}
\end{figure}

We utilize digitized data from the Kodaikanal Observatory (KO) for the present work. The daily Ca-K spectroheliograms obtained at KO with some data gaps for 1907--2007, were digitized with a pixel resolution of 0.86 arcsec and 16 bit readout using a CCD detector of 4K$\times$4K format. \cite{Priyal2019SoPh} have generated a time series of Ca-K images after correcting the limb darkening  and instrumental intensity vignetting. The contrast of the images in the time series was made uniform by applying the ECT methodology \citep{Singh2021ApJ}. The Ca-K features, such as plages, EN, AN, and QN regions, were segregated based on intensity and area threshold values in the different latitude belts with an interval of 10$^{\circ}$ up to 90$^{\circ}$ \citep{Priyal2023ApJ}. In panel (a) of Figure~\ref{fig:fig_1}, we present an example of the analyzed Ca-K image taken on 1936 April 11. The identified plages, EN, AN, and QN, in the binary format, are presented in panels (b), (c), (d), and (e), respectively. These digitized Ca-K data are used from 1907 to 1996 to study the differential rotation. After 1996, there were significant data gaps. Therefore, we excluded the later part (1997--2007) of the data for the present study. Distribution of daily fractional area of plage, EN, AN, and QN  is shown in the four panels of Figure~\ref{fig:fig_frac} for the period 1907--1996.\\

\begin{figure}[!ht]
\plotone{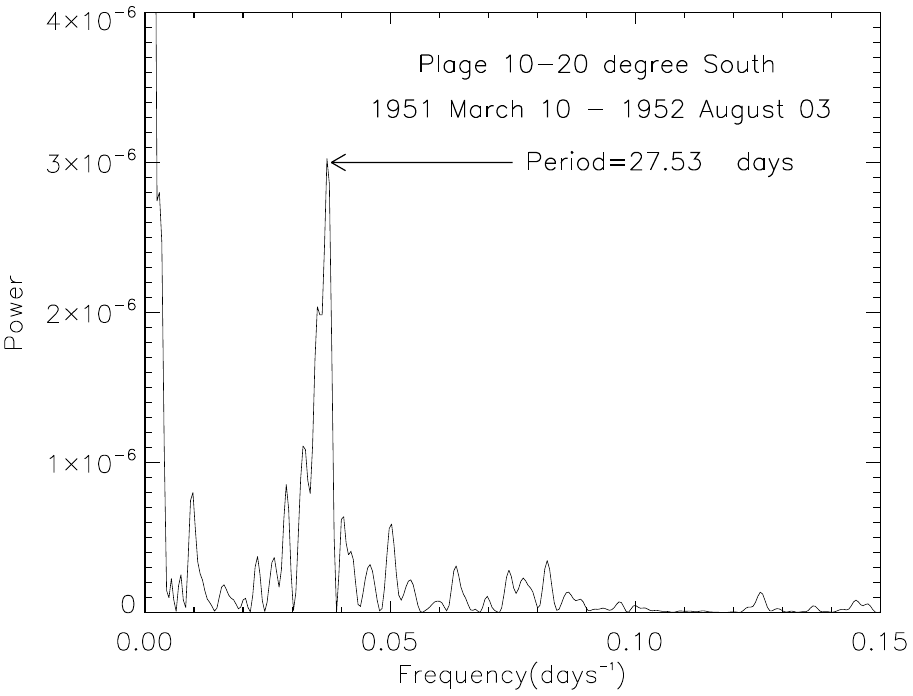}
\caption{A sample FFT power spectra of 512 days of data (from 1951 March 10 to 1952 August 03) is shown for a plage area of 10-20$^{\circ}$ latitude belt.} 
\label{fig:fig_2}
\end{figure}

In the absence of Ca-K data, the plages, EN, AN, and QN areas are interpolated using the IDL subroutine ‘INTERPOL.PRO’ to fill the data gaps and obtain the continuous time series. We have analyzed the data using the power spectral analysis technique. Raw power spectra are obtained through a fast Fourier transform (FFT) for data string of 512 days \citep{Singh1985SoPh}.  The linear trend in the data is removed before performing the FFT. To form successive data string, the data for 256 days are taken common to successive data intervals. The total data length of 32759 days (1907 January 01 to 1996 October 01) thus gave us total 127 epochs. This is done to assure the continuous and at short interval information about the chromospheric rotation period. The power spectra for one epoch encompassing 512 days from 1951 March 10 to 1952 August 03 in the 10--20$^{\circ}$ latitude belt of the southern hemisphere is shown in Figure \ref{fig:fig_2}. The power spectrum shows a prominent peak at $\sim$27 days, corresponding to the solar rotation rate. However, power spectra show double peaks at $\sim$27 days, suggesting the temporal variation of the rotation period over a time scale shorter than the total data length. However, we have selected the most dominant peak only. A similar power spectral analysis is also performed for other epochs, a total of 127 epochs, and the rotation period corresponding to the maximum power is noted.\\

\section{Results and Discussions} \label{sec:result}

We identify the features such as plages, EN, AN, and QN based on intensity threshold and computed areas of each feature, separately,  were grouped into different latitude belts considering the 0$^{\circ}$--10$^{\circ}$, 10$^{\circ}$--20$^{\circ}$, 20$^{\circ}$--30$^{\circ}$, 30$^{\circ}$--40$^{\circ}$, 40$^{\circ}$--50$^{\circ}$, 50$^{\circ}$--60$^{\circ}$, 60$^{\circ}$--70$^{\circ}$, and 70$^{\circ}$--80$^{\circ}$  belts in both the north and south hemispheres. Hereafter, we refer to them by the mean latitude as 5$^{\circ}$, 15$^{\circ}$, 25$^{\circ}$, 35$^{\circ}$, 45$^{\circ}$, 55$^{\circ}$, 65$^{\circ}$, and 75$^{\circ}$ latitude belts.  The derived rotation periods are noted as a function of time for various latitudes. In this way, we have obtained 127 rotation period values for each latitude belt. We have computed the mean and standard deviation of the rotation period for each latitude belt from the pool of data points. In some cases, the power spectrum shows two peaks of similar strength separated by a few days. This can lead to data scatter, particularly at higher latitudes. To address this, we filtered the data by removing values that fell outside one and half times standard deviation. This approach helps concentrate the data points around a central value and reduce the influence of outliers. Finally, we calculated the average and standard deviation of these refined rotation periods to determine the final values for our analysis. Afterward, the rotation period is converted into rotation rate by using the relationship as, Rotation Rate (deg day$^{-1}$)= 360$^{\circ}$/Period (days).

\begin{figure}[!ht]
\plotone{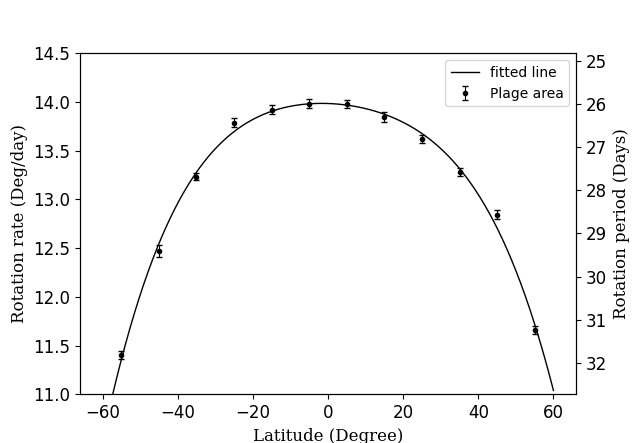}
\caption{Average synodic solar rotation rate measured using the plage area during 1907--1996 as a function of latitude. Error bars are standard error values. Solid line represents the  fourth order polynomial fit as 13.983-0.0013$\phi$-0.0003($\phi$)$^2$+1.54$\times$10$^{-6}$($\phi$)$^3$-1.38$\times$10$^{-7}$($\phi$)$^4$, where $\phi$ is the heliographic latitude.} 
\label{fig:fig_3}
\end{figure}

The average synodic rotation rate measured using plage areas is plotted as a function of heliographic latitude in Figure \ref{fig:fig_3}. Plages are the magnetically active chromospheric structures prominently visible in the Ca II K line (3933.67 $\AA$) and mainly occur up to mid-latitudes. Therefore, its rotation rate is measured only up to $\pm$55$^{\circ}$ latitudes (positive for northern hemisphere and negative for southern hemisphere). The figure shows average rotation rates (represented by dots) for different latitude bands. The solid line represents the best-fit model capturing the overall trend. Error bars indicate the standard error associated with each data point. This model fitting was achieved using a fourth-order polynomial equation. The best-fit to the rotation rate ($\omega$) values is a fourth order polynomial given by the equation, 
\begin{equation}
\omega = A+ B\phi+C\phi^{2}+D\phi^{3}+E\phi^{4}
\label{equation}
\end{equation}

Where, $\phi$ is the heliographic latitude (in degrees). Coefficient \textit{A} represents the rotation rate at the equator. Here, the value of coefficients A, B, C, D, and E are obtained to be 13.983$\pm$0.044, -0.0013$\pm$0.0017, -0.0003$\pm$8.24$\times$10$^{-5}$, 1.54$\times10^{-6}\pm$7.56$\times$10$^{-7}$, and -1.38$\times10^{-7}\pm$2.65$\times$10$^{-8}$, respectively. Using this equation, we have calculated the rotation rate at different latitudes as shown in Table~\ref{tab:rate}. At the equator, the rotation rate is obtained to be 13.983$\pm$0.044 deg day$^{-1}$ for the plage area. As we move towards the higher polar latitudes, the rotation rate values decrease accordingly in both the north and south hemispheres. This trend of rotation rate compares well with previous studies \citep{Lamb2017ApJ,Bertello2020Apj}, but they have used different data sets. \cite{Lamb2017ApJ} used the photospheric magnetic features from high-cadence HMI aboard the SDO, while \cite{Bertello2020Apj} used full-disk Ca II K images from Mount Wilson Observatory. A comparison of the results of rotation rate is presented in \cite{Beck2000SoPh}, \cite{Bertello2020Apj}, and \cite{Jha2021SoPh}.

\begin{figure}[!ht]
\plotone{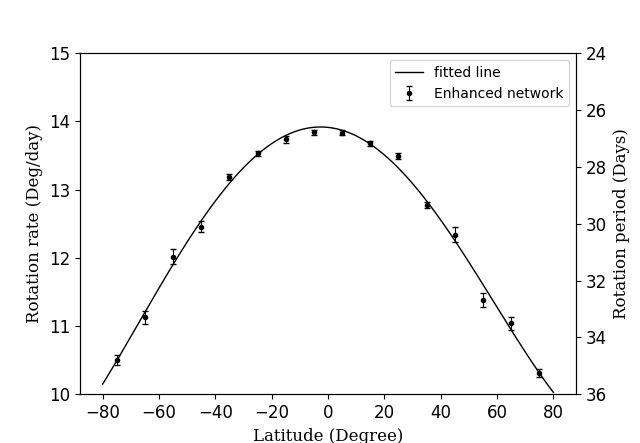}
\caption{Same as Figure \ref{fig:fig_3} but for EN area. Solid line represents the fourth order polynomial fit as 13.912-0.0043$\phi$-0.0008($\phi$)$^2$+5.65$\times$10$^{-7}$($\phi$)$^3$+3.42$\times$10$^{-8}$($\phi$)$^4$, where $\phi$ is the heliographic latitude.} 
\label{fig:fig_4}
\end{figure}

\begin{figure}[!ht]
\plotone{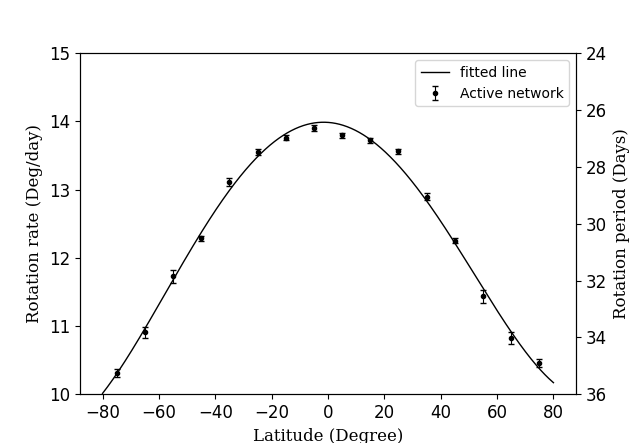}
\caption{Same as Figure \ref{fig:fig_3} but for AN area. Solid line represents the fourth order polynomial equation as 13.985-0.0029$\phi$-0.0009($\phi$)$^2$+6.13$\times$10$^{-7}$($\phi$)$^3$+5.19$\times$10$^{-8}$($\phi$)$^4$, where $\phi$ is the heliographic latitude.} 
\label{fig:fig_5}
\end{figure}

We extended our study to the small-scale network areas. Figures~\ref{fig:fig_4} and \ref{fig:fig_5} present the rotation rate values as a function of heliographic latitude measured by using EN and AN, respectively. We note that EN and AN show a similar variation in rotation rate as for the plage area in Figure~\ref{fig:fig_3}. Here the best-fit model for the rotation rate is a fourth order polynomial given by equation~\ref{equation}. Dots indicate the rotation rate data point at each latitude belt. For EN, the value of coefficients is obtained to be 13.912$\pm$0.054, -0.0043$\pm$0.0015, -0.0008$\pm$5.50$\times$10$^{-5}$, 5.65$\times10^{-7}\pm$3.82$\times$10$^{-7}$, and 3.42$\times10^{-8}\pm$9.80$\times$10$^{-9}$. For AN, the coefficients A, B, C, D, and E are measured as 13.985$\pm$0.053, -0.0029$\pm$0.0015, -0.0009$\pm$5.44$\times$10$^{-5}$, 6.13$\times10^{-7}\pm$3.78$\times$10$^{-7}$, and 5.19$\times10^{-8}\pm$9.70$\times$10$^{-9}$, respectively. The equatorial rotation rate is obtained to be 13.91$\pm$0.05 deg day$^{-1}$ for EN and 13.99$\pm$0.05 deg day$^{-1}$ for AN. These values are comparable to the value obtained for the plage area. 

\begin{figure}[!ht]
\plotone{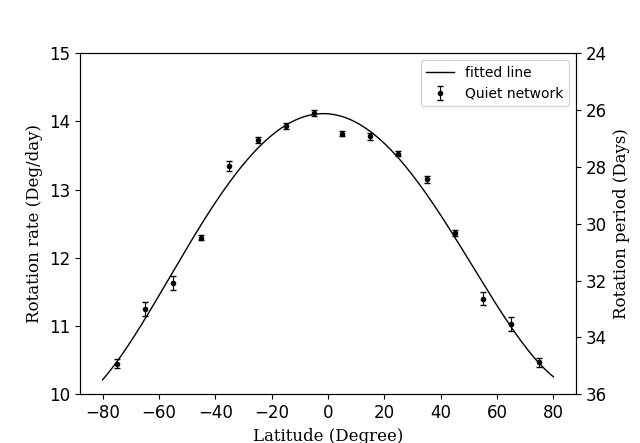}
\caption{Same as Figure \ref{fig:fig_3} but for QN area. Solid line represents the fourth order polynomial equation as 14.111-0.0031$\phi$-0.0009($\phi$)$^2$+5.36$\times$10$^{-7}$($\phi$)$^3$+5.54$\times$10$^{-8}$($\phi$)$^4$, where $\phi$ is the heliographic latitude.} 
\label{fig:fig_6}
\end{figure}

The same analysis is performed on the QN areas. Figure~\ref{fig:fig_6} shows a plot between the rotation rate and the heliographic latitude for the quiet network area. We find that plage, EN, AN, and QN show a similar variation of rotation rate with latitudes. Coefficients A, B, C, D, and E of the fourth-order polynomial fit (given by equation~\ref{equation}) are measured to be 14.111$\pm$0.083, -0.0031$\pm$0.0024, -0.0009$\pm$8.41$\times$10$^{-5}$, 5.36$\times10^{-7}\pm$5.85$\times$10$^{-7}$, and 5.54$\times10^{-8}\pm$1.49$\times$10$^{-8}$, respectively. The rotation rate for QN at the equator is 14.11$\pm$0.08 deg day$^{-1}$ which again compares well with the rotation rate value at the equator obtained using other chromospheric features.

In Table~\ref{tab:comparision}, we compare our results with the previous studies carried out using Doppler shift, magnetic features, sunspot area, and Ca K plage area at the equator and at 40$^{\circ}$ latitude. The rotation rate value at 40$^{\circ}$ latitude is calculated with the help of best-fit equation obtained in the respective studies. We found that our results are comparable with the previous studies. \cite{Komm1993SoPh} measured sidereal rotation rate of small magnetic features (without active regions). They found the equatorial rotation rate as 2.913 $\mu$rad s$^{-1}$ or 14.420 deg day$^{-1}$ and at 40$^{\circ}$ latitude as 13.236 deg day$^{-1}$.

\cite{Gupta1999SoPh} measured rotation rate for three groups of sunspot, sunspot area less than 5 millionth of hemisphere, sunspot area larger than 5, but less than 15 millionth of hemisphere, and sunspot area larger than 15 millionth of hemisphere and found that the smaller sunspots (with small area) rotate faster than the bigger ones (with large area) (see Table~\ref{tab:comparision} for rotation rate values). A similar result was obtained by \cite{Jha2021SoPh}. They have classified sunspots area into two groups: smaller sunspots with sunspot area less than 200 millionth of hemisphere and the bigger sunspots  with sunspot area larger than 400 millionth of hemisphere and then measured the equatorial rotation rate as 14.399 deg day$^{-1}$ for small area group and 14.351 deg day$^{-1}$ for large area group. Therefore, they have concluded that the bigger sunspots group rotates slower than the smaller ones. One  plausible explanation for these results is that sunspots of different sizes are rooted at different depths \citep{Gilman1979ApJ,Beck2000SoPh}. It is found that the large size sunspots suppress the differential rotation in the photosphere \citep{Braj2006SoPh, Li2020ApJ}. The smaller rotation rates of larger sunspot thus may indicate the deeper anchoring depth of the parent ﬂux tubes \citep{Livingston2006SoPh}. A detailed and comprehensive analysis of the physical mechanisms underlying the differential rotation rates of different sized sunspots is presented by \cite{Wan2023ApJ}.

\begin{figure}[!ht]
\plotone{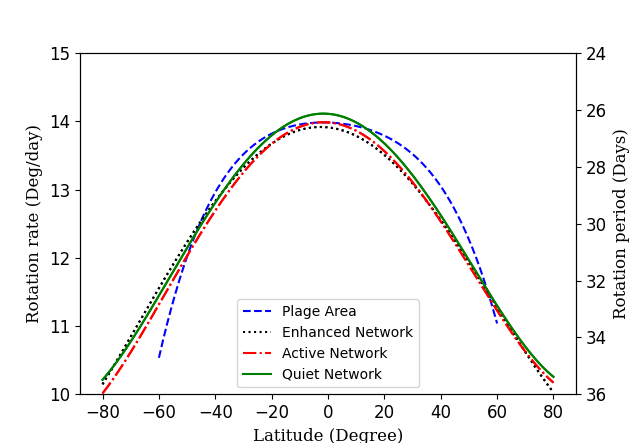}
\caption{Comparision of differential rotation rate profiles obtained for plage area (blue dashed line), EN (black dotted line), AN (red dash-dotted line), and QN (green solid line).} 
\label{fig:fig_compare}
\end{figure}

A plot for the comparison of rotation rates of plage area and small-scale networks is presented in Figure~\ref{fig:fig_compare}. Rotation rate of plages are slightly higher than that of small-scale networks at mid-latitudes, slightly lower at higher latitudes and comparable to small-scale network near the equator (Table~\ref{tab:rate}). However, the difference in the rotation rate of two populations is small. Therefore, we can infer that the small scale chromospheric features observed in Ca-K line show a similar rotation rate as large structures such as plages. This indicates that large and small-scale features have the foot points at the same layer of the Sun. \cite{Hale1908ApJ} obtained that the rotation rate of plages are consistent with that of sunspots or surface magnetic field. The rotation rates obtained by the quiet background network emission pattern in the Ca II K line are in close agreement with those of small sunspots \citep{Schroeter1978SoPh}. \cite{Belvedere1977ApJ} investigated that the large evolved chromospheric plages rotate
more rigidly than the smaller ones. However, \cite{Antonucci1979SoPh} found that the chromospheric features with sizes 24--300 $\times$ 10$^{3}$ km, such as network elements and active regions, exhibit the same differential rotation. Recent studies by \cite{Li2020ApJ} and \cite{Li2023MNRAS} reported that the quiet chromosphere rotates at the same rate as the small-scale magnetic elements do. In the chromosphere, in contrast to photosphere, the rotation rate is strengthened by large-scale magnetic field \citep{Li2013ApJS,Wan2023MNRAS}. Further, it has been suggested that magnetic elements in quiet regions rotate marginally faster than sunspots in active regions \citep{Xiang2014AJ, Xu2016ApJ}. Therefore, the driving force for the chromosphere to rotate in this way is expected from the solar interior, not from the photosphere \citep{Li2020ApJ}.

\section{Conclusions} \label{sec:conclusion}

In this study, the digitized data of four chromospheric features \textit{viz} plage, AN, EN, and QN area obtained from Kodaikanal Observatory for the period 1907--1996 are used to investigate the differential rotation at different latitude belt from 0 to 80 degree with a step size of 10$^{\circ}$ in both  hemispheres. We find that plage areas as well as chromospheric networks are showing the pattern of solar differential rotation. Even QNs (most fainter structures) are found to show the differential rotation profile. From this, we concluded that chromospheric features like plage, EN, AN, and even QN areas are reliable parameters for the study of differential rotation of the solar chromosphere. Moreover, this time, we have used the data of higher polar latitudes up to $\pm$80$^{\circ}$ for the study of differential rotation. We observed that rotation rate presents a decreasing trend from equator to poles regardless of solar structure type used to measure the rotation rate. The equatorial rotation rate for plage, EN, AN, and QN area are estimated to be 13.983$\pm$0.044 deg day$^{-1}$, 13.912$\pm$0.054 deg day$^{-1}$, 13.985$\pm$0.053 deg day$^{-1}$, and 14.111$\pm$0.083 deg day$^{-1}$, respectively. The results of this study will help to understand the differential rotation and so the dynamo theory of the Sun. In this way, further analyses can be done in future by using the data of other observatories and by other methods; and a comparision of such studies will definitely enhance our understanding toward the important phenomena of solar differential rotation.   

\begin{acknowledgments}
We thank the anonymous referee for providing valuable comments and suggestions on the paper. We would like to thank all the observers at Kodaikanal Observatory who made the observations, maintained the data at the observatory, and digitized all the images. The data digitization was supervised by Jagdev Singh and supported by F. Gabriel and P U. Kamath. 
\end{acknowledgments}

\typeout{}
\bibliography{diff}
\bibliographystyle{aasjournal}

\begin{deluxetable*}{ccccc}
\tablenum{1}
\tablecaption{Differential rotation rate measured using plage, EN, AN, and QN area.\label{tab:rate}}
\tablewidth{0pt}
\tablehead{
\colhead{Latitude} & \nocolhead{Common} & \colhead{Rotation rate (Rotation period)} & \nocolhead{Common} &
\nocolhead{Common}\\
\colhead{(Degree)} & \nocolhead{Common} & \colhead{deg day$^{-1}$ (days)} & \nocolhead{Common} &
\nocolhead{Common}\\
\colhead{} & \colhead{Plage} & \colhead{EN} & \colhead{AN} &
\colhead{QN}
}
\decimalcolnumbers
\startdata
-80 &-  & 10.146 (35.481) & 10.012 (35.953) & 10.211 (35.252)\\
-70 &-  & 10.843 (33.201) & 10.618 (33.904) & 10.772 (33.418)\\
-60 & 10.534 (34.173) & 11.554 (31.155) & 11.315 (31.813) & 11.445 (31.453)\\
-50 & 12.018 (29.954) & 12.231 (29.432) & 12.028 (29.928) & 12.147 (29.636)\\
-40 & 12.960 (27.777) & 12.831 (28.056) & 12.691 (28.365) & 12.808 (28.106)\\
-30 & 13.518 (26.629) & 13.320 (27.026) & 13.252 (27.164) & 13.371 (26.922)\\
-20 & 13.819 (26.049) & 13.673 (26.327) & 13.671 (26.332) & 13.794 (26.096)\\
-10 & 13.955 (25.797) & 13.873 (25.947) & 13.920 (25.861) & 14.046 (25.628)\\
 0 & 13.983 (25.744) & 13.912 (25.876) & 13.985 (25.741) & 14.111 (25.511)\\
10 & 13.931 (25.841)  & 13.787 (26.110)& 13.862 (25.968) & 13.984 (25.742)\\
20 & 13.790 (26.105)  & 13.508 (26.650)& 13.563 (26.542) & 13.676 (26.321)\\
30 & 13.520 (26.625) & 13.089 (27.503) & 13.109 (27.462) & 13.211 (27.249)\\
40 & 13.049 (27.588) & 12.554 (28.675) & 12.534 (28.720) & 12.624 (28.516)\\
50 & 12.268 (29.344) & 11.936 (30.160) & 11.887 (30.283) & 11.965 (30.086)\\
60 & 11.038 (32.613) & 11.275 (31.928) & 11.227 (32.063) & 11.298 (31.863)\\
70 & -  & 10.619 (33.900) & 10.627 (33.875) & 10.698 (33.649)\\
80  & - & 10.026 (35.906) & 10.170 (35.397) & 10.255 (35.101)\\
\enddata
\end{deluxetable*}

\begin{deluxetable*}{cccc}
\tablenum{2}
\tablecaption{Comparision of rotation rate measured for different features.\label{tab:comparision}}
\tablewidth{0pt}
\tablehead{
\colhead{Author} & \colhead{Features} & \colhead{Rotation rate (deg day$^{-1}$)} & \nocolhead{}\\
\nocolhead{} & \nocolhead{} & \colhead{at equator} & \colhead{at 40$^{\circ}$ latitude}\\
}
\decimalcolnumbers
\startdata
\cite{Howard1970SoPh} & Doppler Shift & 13.761 & 12.675\\
\cite{Komm1993SoPh} & small magnetic features & 14.420 & 13.236\\
\cite{Gupta1999SoPh}& Sunspot area$<$5$\mu$Hem & 14.491 & 13.313\\
\cite{Gupta1999SoPh}& 5$\mu$Hem$<$Sunspot area$<$15$\mu$Hem&14.380&13.206\\
\cite{Gupta1999SoPh}& Sunspot area$>$15$\mu$Hem & 14.279 & 13.109\\
\cite{Lamb2017ApJ} & magnetic feature & 14.296 & 13.086\\
\cite{Bertello2020Apj} & Ca II K plage & 14.286 & 13.024\\
\cite{Jha2021SoPh} & Sunspot area & 14.381 & 13.257 \\
\cite{Wan2022ApJ} & Ca II K plage & 13.496 & 12.960\\
Present work  & Plage & 13.983 & 12.960\\
Present work  & EN & 13.912 & 12.831\\
Present work  & AN & 13.985 & 12.691\\
Present work  & QN & 14.111 & 12.808\\
\enddata
\end{deluxetable*}

\end{document}